\documentclass[12pt, a4paper]{article}

\usepackage{graphicx}
\usepackage{bmpsize}
\usepackage{amsmath}
\usepackage{amsfonts}
\usepackage{amssymb}
\usepackage{latexsym}
\usepackage{graphicx}
\usepackage{color}
\usepackage{mathrsfs}
\usepackage{slashed}
\usepackage{soul}
\usepackage{array}
\usepackage{cite}
\usepackage{placeins}
\usepackage{hyperref}

\usepackage{enumerate} 
\usepackage{multirow}
\usepackage[table]{xcolor}
\usepackage{colortbl}
\usepackage{tikz-feynman}
\tikzfeynmanset{compat=1.0.0}

\usepackage{xfrac}

\usepackage{subcaption}

\usepackage{float}
\usepackage{braket}
\usepackage
[
version=4,
math-greek=default,
text-greek=default,
]{mhchem
}
\usepackage{booktabs}  

\usepackage[text={7in,10in}]{geometry}

\usepackage{comment}
\includecomment{toexclude} 
\excludecomment{addsections} 

\usepackage[
  output-decimal-marker=.,
  separate-uncertainty=true,
  per-mode=symbol-or-fraction,
]{siunitx}
\DeclareSIUnit\parsec{pc}
\DeclareSIUnit\erg{erg}

\numberwithin{equation}{section}

 \def\be   {\begin{equation}}   \def\ee   {\end{equation}}
       \def\ea   {\end{array}}
 \def\bea  {\begin{eqnarray}}   \def\eea  {\end{eqnarray}}
 \def\bean {\begin{eqnarray*}}  \def\eean {\end{eqnarray*}}

\newcommand{\half}{\frac{1}{2}}
\newcommand{\lag}{\mathcal{L}}
\newcommand{\hc}{\text{h.c.}}
\newcommand{\ab}{{\alpha\beta}}
\newcommand{\inbr}[1]{\left(#1\right)}

\begin{document}
\hfill\textit{DO-TH 18/23}


\begin{center}
{\Huge
Massive Majorons and Constraints on the Majoron-Neutrino Coupling
}
\\ [2.5cm]
{\large{\textsc{ 
Tim Brune\footnote{\textsl{tim.brune@tu-dortmund.de}}
}}}
{\large{\textsc{ 
Heinrich P\"as\footnote{\textsl{heinrich.paes@tu-dortmund.de}}
}}}
\\[1cm]

\large{\textit{
Fakult\"at f\"ur Physik, Technische Universit\"at Dortmund,\\
44221 Dortmund, Germany
}}
\\ [2 cm]
{ \large{\textrm{
Abstract 
}}}
\\ [1.5cm]
\end{center}
We revisit a singlet Majoron model in which neutrino masses arise from the spontaneous violation of lepton number. If the Majoron obtains a mass of order $\si{MeV}$, it can play the role of dark matter. We discuss constraints on the couplings of the massive Majoron with masses of order $\si{MeV}$ to neutrinos from supernova data. In the dense supernova core, Majoron-emitting neutrino annihilations are allowed and can change the signal of a supernova. Based on the observation of SN1987A, we exclude a large range of couplings from the luminosity and the deleptonization arguments, taking the effect of the background medium into account. If the Majoron mass does not exceed the Q-value of the experiment, the neutrino-Majoron couplings allow for neutrinoless double beta decay with Majoron emission. We derive constraints on the couplings for a Majoron mass of order $\si{MeV}$ based on the phase space suppression and the diminishing signal-to-background ratio due to the Majoron mass. The combination of constraints from astrophysics and laboratory experiments excludes a large range of neutrino-Majoron couplings in the mass range of interest for Majoron dark matter, \textcolor{black}{where they complement existing cosmological bounds from dark matter stability and the effects of a decaying Majoron on the cosmic microwave background anisotropy spectrum. }

\def\thefootnote{\arabic{footnote}}
\setcounter{footnote}{0}
\pagestyle{empty}

\newpage
\pagestyle{plain}
\setcounter{page}{1}

\section{Introduction} 
\label{sec:intro}
The observation of neutrino oscillations \cite{sno, kamland, superkamiokande} gives evidence to at least two nonvanishing neutrino masses much smaller than the masses of the other standard model (SM) particles. As the SM still lacks an explanation for neutrino masses and their smallness, a large variety of neutrino mass generating mechanisms has been explored over the past years with the most popular one being the seesaw mechanism \cite{minkowski}. In the seesaw mechanism, heavy right-handed neutrinos suppress the masses of the left-handed neutrinos and thus offer a natural explanation for the smallness of the neutrino mass. This mechanism requires neutrinos to be Majorana particles and consequently leads to a violation of baryon-lepton number $U(1)_{B-L}$ by two units. Assuming that $U(1)_{B-L}$ is a global symmetry and that the symmetry breaking occurs spontaneously, a massless Goldstone boson, called the Majoron, will be generated \cite{chikashige, schechtervalle, gelmini, georgi}. Models with massless (or very light) Majorons have been studied extensively in the literature, where the Majoron was originally either \textcolor{black}{a singlet \cite{chikashige} or  part of a doublet or triplet} \cite{gelmini, georgi}. However, the last two options are ruled out due to contributions to the invisible $Z$-width via decays of the $Z$ boson to the Majoron and its scalar partner, equivalent to one-half or two extra neutrino species \cite{zwidth}.\\
Currently, a major part of research in particle physics is dedicated to the ongoing search for a dark matter (DM) particle. The Majoron as a DM particle has already been discussed in \cite{majorondm1, majorondm2} and as the search for DM continues to be unsuccessful, the interest in Majoron models is recently reviving. 
An appealing feature of Majoron models with respect to DM is the suppression of the couplings of the Majoron to SM fermions by the seesaw-scale, rendering it stable on cosmological timescales. If the Majoron acquires a mass and becomes a pseudo-Goldstone boson, it will be a viable DM candidate \cite{main}. \\
Constraints on the couplings of the Majoron to neutrinos can be derived from astrophysics as well as from laboratory experiments. First, the Majoron can have a significant impact on the process of explosion and cooling of a supernova (SN). Second, constraints can be derived from laboratory experiments searching for neutrinoless double beta decay with Majoron emission. While the constraints for the case of a massless Majoron have been discussed in great detail (see for example \cite{zwidth, farzan, kachel, tomas, burgess} and references therein), models including a massive Majoron have rarely been considered.\footnote{Recently, constraints on a massive Majoron from SN data have been derived in \cite{heurtierzhang} and constraints from double beta decay have been discussed in \cite{blum}. In contrast to \cite{heurtierzhang, blum}, we include the impact of the effective potentials and data from other experiments searching for neutrinoless double beta decay with Majoron emission. Moreover, in \cite{carone}, constraints on a massive vector Majoron have been derived.} In this work, we aim to perform a dedicated analysis of the constraints on the neutrino-Majoron couplings from SN data and neutrinoless double beta decay \textcolor{black}{for Majorons in the $\si{MeV}$ mass range}. \\
This paper is organized as follows. In Sec. \ref{sec:interactions}, we discuss neutrino-Majoron interactions in vacuum and in matter. In Sec. \ref{sec:dm}, we briefly discuss a mechanism to generate a Majoron mass and the possibility of Majoron DM. In Sec. \ref{sec:snconstraints} and \ref{sec:0nubbconstraints}, we derive bounds on the neutrino-Majoron couplings from SN data and neutrinoless double beta decay with Majoron emission which we compare in Sec. \ref{sec:comp}. We conclude in Sec. \ref{sec:conclusion}.

\section{Majoron Interactions}
\label{sec:interactions}
The Lagrangian coupling neutrinos to the Majoron can be in general written as 
\begin{align}
  \lag_\text{int} \propto \sum_{ij}g_{ij} \nu_i \gamma_5 J \bar \nu_j 
\end{align}
where in vacuum, the Majoron couples diagonally to the neutrino mass eigenstates, i.e. \textcolor{black}{$g_{ij}=\delta_{ij}g_i\propto\frac{m_i}{f}$. }
In the presence of a background medium, the neutrino-Majoron interactions are modified, which will be discussed in the following. \\
In general, when propagating in a medium, flavor neutrinos interact with the background medium coherently via charged-current (CC) and neutral-current (NC) interactions. 
This gives rise to effective potentials that shift the energy of the neutrinos and therefore change the evolution equation. An example for a medium where the neutrino interactions with matter have to be taken into account is the core of a SN. The corresponding effective potentials are given by
\begin{align}
  V_C &= \sqrt{2}G_Fn_B(Y_e+Y_{\nu_e})\;, \\
  V_N &= \sqrt{2}G_Fn_B(-\frac{1}{2}Y_N + Y_{\nu_e})\;,
  \label{eq:effpot}
\end{align}
where $n_B$ is the baryon density, $G_F$ is the Fermi coupling constant, and the particle number fraction $Y_i$ is defined as
\begin{align}
  Y_i = \frac{n_i-n_{\bar{i}}}{n_B}\;.
\end{align}
The background medium in a SN core consists mostly of electrons $e$, protons $p$ and neutrons $n$. Therefore, electron neutrinos $\nu_e$ can have CC and NC interactions with the background medium and their effective potential is given by 
\begin{align}
V_e^{(h)} = V_C + V_N = -h\sqrt{2}G_Fn_B(Y_e + 2Y_{\nu_e} - \frac{1}{2}Y_N)\;,
\label{eq:pote}
\end{align}
where $h = \pm 1$ is the helicity of the respective neutrino.
Muon and tau neutrinos $\nu_{\mu,\tau}$ can only have NC interactions and their effective potential is consequently given by 
\begin{align}
V_{\mu,\tau}^{(h)}  = -hV_N\;. 
\label{eq:potx}
\end{align}
The Hamiltonian describing the neutrino evolution has to be extended by a term that takes the flavor-dependent energy shift in matter into account. Therefore, the mass eigenstates $\ket{\nu_i}$ will no longer be eigenstates in matter and it will be necessary to introduce a third type of eigenstate, the medium eigenstate $\ket{\tilde\nu_i}$. As shown in App. \ref{app:int}, in dense media, the medium eigenstate $\ket{\tilde\nu_i}$ can be approximated as the weak state $\ket{\nu_\alpha}$ with medium energy eigenvalues 
\begin{align}
  E^{(h)} = p + V^{(h)}\;
\end{align}
 and nondiagonal neutrino-Majoron couplings in medium 
\begin{align}
   \tilde g_{fm} = g_\ab = U_{\alpha i}^* g_{ij} U_{\beta j}\;.
\end{align}
 Thus, in medium, the Majoron couples effectively to the neutrino flavour eigenstates.

\section{\textcolor{black}{Massive Majorons as Dark Matter}}
\label{sec:dm}
In this section, a mechanism to generate a nonvanishing Majoron mass is discussed. We stress that the constraints on the neutrino-Majoron couplings derived in \ref{sec:snconstraints} and \ref{sec:0nubbconstraints} do not depend on the mass-generating mechanism and other possibilities exist, see for example \cite{frigerio}. \\
A Majoron mass can be generated by explicitly breaking the global $U(1)_{B-L}$ symmetry via a radiatively induced term \cite{main, frigerio},
\begin{align}
  \lag_H &= \lambda_h \sigma^2 H^\dagger H + \hc\;
  \label{eq:portal}
\end{align}
After spontaneous symmetry breaking at the seesaw-scale $f$ and electroweak symmetry breaking at the scale $v$, a Majoron mass $m_J$ is generated via
\begin{align}
      \lag_H = -\half m_J^2 J^2(1+ \frac{h}{v})^2 \;,
\end{align}
where the mass of the Majoron is directly proportional to the VEV of the Higgs, $m_J^2 = \lambda_h v^2$. 
In order for the Majoron to account for DM, the Majoron relic density $\Omega_J$ has to coincide with the DM relic density. For simplicity, the only Majoron production mechanism considered in the following discussion is the Higgs decay $h \to J J $.\footnote{It has been shown in \cite{frigerio} that the Higgs decay dominates the Majoron-producing scattering processes, assuming small Yukawa couplings of the heavy neutrinos to $J$ in order to neglect Majoron production at the seesaw-scale.}
The corresponding decay rate is given by \cite{frigerio}
\begin{align}
  \Gamma(h\to JJ) = \frac{1}{16\pi}\lambda_h^2 \frac{v^2}{m_h^2}\sqrt{1-4\frac{m_J^2}{m_h^2}} 
  \label{eq:higgsdecay}
\end{align}
and depends only on $m_J$.
There exist two well-known production mechanisms for DM, known as the freeze-out and the freeze-in mechanisms.
The scenario of the freeze-out mechanism (for details, see \cite{frigerio, hall}) is ruled out in the mass range of interest by constraints from direct detection or $h \to \text{invisible}$, as shown in shown in \cite{cline}. However, the freeze-in mechanism is capable of producing the correct Majoron relic density, as we will discuss in the following. \\ 
In the freeze-in mechanism \cite{hall}, the initial abundance of the DM particle is negligible with respect to those of the SM particles after reheating. The DM particle is produced via the decay of a heavier particle $X$. If the decay rate is small enough, it will never thermalize. As the temperature reaches $T \approx m_{X}$, the DM density reaches a plateau due to the Boltzmann suppression of the heavy particle. 
An interesting feature of the freeze-in mechanism is that the relic abundance of DM increases with the coupling to the heavy particle. 
In the freeze-in scenario, the Majoron relic density is given by \cite{hall, main}
\begin{align}
  \Omega_J h^2 \approx 2\frac{1.09\times 10^{27}}{g_*^s \sqrt{g_*^\rho}}\frac{m_J \Gamma(h\to JJ)}{m_h^2}\;.
  \label{eq:abundance}
\end{align}
Using \eqref{eq:higgsdecay}, the only free parameter in \eqref{eq:abundance} is 
 $m_J$ and $\Omega_{J}h_0^2 \approx 0.12$ can be fulfilled for 
\begin{align}
  m_J \approx \SI{2.8}{MeV}\;,
  \label{eq:dmmj}
\end{align}
which translates to a coupling
\begin{align}
  \lambda_h \approx \num{1.3 e-10}\;.
\end{align}
Note that in our discussion, we assumed the Majoron to be the only DM particle. If other DM particles exist, the Majoron has to account only for a fraction of the DM relic density, translating to $m_J \lesssim \SI{2.8}{MeV}$, i.e. \eqref{eq:dmmj} is an upper bound on the Majoron mass.\\

\textcolor{black}{ 
A stringent bound on the neutrino-Majoron coupling can be derived from DM stability which requires the lifetime of the Majoron to exceed the age of the universe. As has been shown in \cite{main}, this can easily be achieved for $m_J \approx\SI{1}{MeV}$ by assuming $f \geq \SI{e9}{GeV}$. In the case of normal ordering, i.e. $m_i \lesssim\SI{e-2}{eV}$ from neutrino oscillations, this translates to a strong constraint on the neutrino-Majoron coupling, $g < 10^{-20}$, in order for the Majoron to be DM. Moreover, the anisotropies of the cosmic microwave background (CMB) can be used to derive constraints on the lifetime of the Majoron \cite{cmb1, cmb2}, resulting in a similar constraint on the neutrino-Majoron couplings, approximately $g < 10^{-20}$. }

\section{Supernova Core-Collapse with Majorons}
\label{sec:snconstraints}
In this section, constraints on the neutrino-Majoron couplings $g_\ab$ for a Majoron mass range $\SI{0.1}{\mega\electronvolt} \lesssim m_J \lesssim \SI{1}{\giga\electronvolt}$ are derived based on SN data. For simplicity, we assume that only one neutrino-Majoron coupling constant $g_\ab$ is nonzero. \\
In the following, the inner core radius is approximated to be $R_C \approx \SI{10}{\kilo\metre}$ with a temperature of $T \approx \SI{30}{\mega\electronvolt}$. The abundance of electron neutrinos in a SN core is extremely high, thus they have a chemical potential of $\mu_{\nu_e} \approx \SI{200}{\mega\electronvolt}$, while the chemical potential of the  electron antineutrinos is given by $\mu_{\bar\nu_e} \approx -\SI{200}{\mega\electronvolt}$. In the first approximation, the chemical potentials of $\overset{\scriptscriptstyle(-)}{\nu}_{\mu,\tau}$ vanish, $\mu(\overset{\scriptscriptstyle(-)}{\nu}_{\mu,\tau}) \approx 0$ \cite{burrows}. In the core, the effective potentials are of order 
\begin{align}
  |V_{e}| &\approx \mathcal{O}(\SI{1}{\electronvolt}) \label{eq:ve}  \;, \\
  |V_{\mu,\tau}| &\approx \mathcal{O}(\SI{10}{\electronvolt}) \label{eq:vx}\;.
\end{align}
There are three different constraints on the neutrino-Majoron couplings $g_\ab$ from SN data to be examined in the following. Our approaches concerning the ``luminosity constraint'' and the ``trapping constraint'' in Sec. \ref{sec:luminosity} and Sec. \ref{sec:trapping}, respectively, follow \cite{heurtierzhang}, with the difference of explicitly including the contribution of the effective potentials in \ref{sec:luminosity}. 
A different approach to obtain the ``deleptonization constraint'' is presented in Sec. \ref{sec:deleptonization}.

\subsection{Constraints from Majoron Luminosity}
\label{sec:luminosity}
The predicted amount of binding energy released in a SN explosion is compatible with the neutrino signal measured from SN1987A \cite{kamio1, kamio2, baksan1, baksan2, imb}.
Consequently, introducing an additional particle, in this case the Majoron, must not alter the signal significantly to be in agreement with experiment. Therefore, the impact of the process $\overset{\scriptscriptstyle(-)}{\nu}\overset{\scriptscriptstyle(-)}{\nu}\to J$ has to be considered, as it can lead to an additional energy depletion that changes the neutrino signal. 
Constraints on the neutrino-Majoron couplings $g_\ab$ are derived under the terms that the  Majoron luminosity does not exceed the total neutrino luminosity within one second after the explosion, $L_J \approx \SI{5e52}{\erg\per\second}$.\\
The luminosity of the inverse Majoron decay $ \nu\nu\to J $ is given by
\begin{align}
\begin{split}
L_J(\overset{\scriptscriptstyle(-)}{\nu}_\alpha\overset{\scriptscriptstyle(-)}{\nu}_\beta\to J) = &\frac{4}{3} \pi R_C^3 DQ(\overset{\scriptscriptstyle(-)}{\nu}_\alpha\overset{\scriptscriptstyle(-)}{\nu}_\beta\to J) \;.
\end{split}
\end{align}
The decay factor
\begin{align}
  D = e^{-\Gamma(J\to\overset{\scriptscriptstyle(-)}{\nu}_\alpha\overset{\scriptscriptstyle(-)}{\nu}_\beta) R_C}
\end{align}
takes into account that the Majoron can decay back to neutrinos inside the core which would prevent an exotic energy depletion. Taking the effective potentials into account, the decay width is given by 
\begin{align}
\begin{split}
  \Gamma(J \to \overset{\scriptscriptstyle(-)}{\nu}_\alpha\overset{\scriptscriptstyle(-)}{\nu}_\beta) &= \int_0^{p_J}\!\!\frac{|g_{\alpha\beta}|^2}{8\pi}\frac{1}{p_J}\left[m_J^2\left(\frac{1}{E_J} - \frac{{\color{black} V_\alpha+ V_\beta}}{2p_\beta(E_J-p_\beta)}  \right) -  \left( V_\alpha+ V_\beta\right)  \right] dp_\beta \;.
  \label{eq:majorondecay}
\end{split}
\end{align}
For $m_J \geq \SI{1}{\kilo\electronvolt}$, the contribution of the effective potentials can be neglected.\footnote{Due to numerical instabilities, we only present constraints for $m_J \geq \SI{0.1}{\mega\electronvolt}$, which still covers the mass range of interest.} Therefore, the decay width reduces to 
\begin{align}
\Gamma(J\to\overset{\scriptscriptstyle(-)}{\nu}_\alpha\overset{\scriptscriptstyle(-)}{\nu}_\beta) \approx \frac{|g_{\alpha\beta}|^2}{8\pi}\frac{m_j^2}{E_J} \;.
\end{align}
The energy emission rate \cite{choisantamaria, dent, heurtierzhang} for the process $\overset{\scriptscriptstyle(-)}{\nu}\overset{\scriptscriptstyle(-)}{\nu}\to J$ is given by
\begin{align}
  Q(\overset{\scriptscriptstyle(-)}{\nu}_\alpha\overset{\scriptscriptstyle(-)}{\nu}_\beta\to J) &= \int \!\!{d}\Pi_\alpha {d}\Pi_\beta {d}\Pi_J E_J|\mathcal{M}|^2F_Sf_\alpha f_\beta(2\pi)^4\delta^{(4)}(P_\alpha + P_\beta - P_J)\;,
  \label{eq:q}
\end{align}
where the symmetry factor $F_S = \frac{1}{1+\delta_{\alpha\beta}}$
takes identical particles in the initial state into account and the Fermi-Dirac distribution $f_\alpha$ is approximated as a Maxwell-Boltzmann distribution, $f_\alpha \approx \frac{1}{\exp{\frac{E_\alpha-\mu_\alpha}{T}}}\;$. 
We find $Q \to 0$ as $m_J \gtrsim T$, which is expected since the Majoron production is suppressed by $e^{-\frac{m_J}{T}}$ and thus extremely ineffective for Majoron masses $m_J \gtrsim T$. 
The only free parameters in \eqref{eq:q} are $m_J$ and $g_{\alpha\beta}$, thus we evaluate bounds on $|g_\ab|$ for Majoron masses $\SI{0.1}{\mega\electronvolt} < m_J < \SI{1}{\giga\electronvolt}$, demanding $L_J < \SI{5e52}{\erg\per\second}$. For a rough approximation, \eqref{eq:ve} and \eqref{eq:vx} are used. The constraints are shown in Fig.  \ref{fig:snconstraints}, where the colored regions are excluded. 
A large range of couplings $|g_{ee}|$ and $|g_{e\alpha}|\,,\, (\alpha = \mu,\,\tau)$ is excluded, while the constraints on $|g_{\alpha\alpha}|$ are comparably weak. This can be traced back to the high abundance of electron neutrinos and the low abundance of muon and tau neutrinos, resulting in neutrino flavor dependent Majoron luminosities as 
$L_J(\nu_e\nu_e\to J)~>~L_J(\nu_e\nu_\alpha \to J)~>~L_J(\nu_\alpha\nu_\alpha \to J)\;.$
At $m_J \approx \SI{2.8}{\mega\electronvolt}$, the mass of interest regarding Majoron DM produced via freeze-in, constraints on $|g_{ee}|$ and $|g_{e\alpha}|\,,\, (\alpha = \mu,\,\tau)$ are derived, while $|g_{\alpha\alpha}|$ is not constrained at this certain Majoron mass. \\
We stress that the constraints suffer from very poor experimental data from SN1987A. As discussed in \cite{heurtierzhang}, the detection of a future SN at a distance of order $\SI{1}{\kilo\parsec}$ could reinforce the constraints on $g_\ab$ and would allow us to probe couplings $|g_\ab|$ up to $\num{e-13}$.

\subsection{Deleptonization Constraints}
\label{sec:deleptonization}
The strength of the SN bounce shock depends on the trapped lepton fraction during the infall stage, 
 $ Y_L = Y_e + Y_{\nu_e}\;$,
which has to be larger than $Y_L^\text{Bounce} \approx 0.375$ at the time of the core bounce in order to allow for a successful explosion\cite{bruennD, baron,berezhianismirnov} . \\
The inverse Majoron decay $\nu_e\nu_\alpha\to J$ changes $\Delta L_e$ by one ($\alpha \neq e$) or two ($\alpha = e$) units and if the neutrino-Majoron coupling is too large, it could prevent a successful explosion. \\
The deleptonization rate for the $\Delta L_e=2$ process $\nu_e\nu_e \to J$ can be calculated in terms of the Boltzmann equation \cite{strumia}
\begin{align}
\dot Y_L &= -2\frac{1}{n_B}\gamma^{eq}(\nu_e\nu_e\to J) \;,
\label{eq:delbol}
\end{align}
where the factor of $2$ takes into account that the process violates electron lepton number by two units and $\gamma^{eq}$ is the thermal rate, given by 
\begin{align}
\gamma^{eq}(\nu_e\nu_e \to J) &= \int\!\! {d}\Pi_\alpha {d}\Pi_\beta {d}\Pi_J |\mathcal{M}|^2F_SDf_\alpha f_\beta(2\pi)^4\delta(P_\alpha + P_\beta - P_J) \;,
\label{eq:thermalrate}
\end{align}
which differs from \eqref{eq:q} only by a factor of $E_J$. We suppose $\nu_e\nu_e \to J$ is the only process violating electron lepton number, thus $n_B\dot Y_L = \dot n_{\nu_e}$. Additionally, the number density of the Majoron is neglected and neutrinos are taken to be in thermal equilibrium. The number density of electron antineutrinos is small compared to the number density of electron neutrinos. Therefore, the impact of the process $\bar\nu_e \bar\nu _e \to J$ has been neglected.\\
We solve the differential equation numerically, with the initial condition 
$Y_L^\text{Infall}(\SI{e12}{\gram\per\centi\meter^3}) \approx 0.75\; $
at the time when neutrinos become trapped.
Constraints on $|g_{ee}|$ for the Majoron mass range of interest are derived under the condition  
$Y_L^\text{Bounce}(\SI{3e14}{\gram\per\centi\meter^3}) \geq 0.375 $
at the time of the core bounce \cite{delcond}. The constraints are visualized in Fig.  \ref{fig:snconstraints}, where the colored regions are excluded. The deleptonization constraints for $|g_{ee}|$ are less stringent than the luminosity constraints on $|g_{ee}|$ and do not exclude an additional range of couplings.  
At $m_J \leq \SI{100}{\mega\electronvolt}$, a small range of couplings $|g_{ee}|$ is not excluded due to the deleptonization constraints, however, the luminosity constraints on $|g_{ee}|$ are still valid in this region.  
On the other hand, using the same method as above, no deleptonization constraints for $|g_{e\alpha}|\,,\, (\alpha = \mu,\,\tau)$ could be derived. First, the abundance of nonelectron neutrinos during the infall stage is much lower than the abundance of electron neutrinos and, second, the process $\nu_e\nu_\alpha \to J$ violates electron lepton number only by one unit. Consequently, the deleptonization constraints for $|g_{e\alpha}|$ are expected to be significantly less stringent compared to the constraints for $|g_{ee}|$.\\ 
We want to stress the strong dependence of the deleptonization constraints on the explosion mechanism, which is not yet well understood, and the numerical modulation. Moreover, the density profiles used in this section are taken from SN simulations without Majoron processes, thus including the Majoron in SN simulations could improve the deleptonization constraints. 

\subsection{Majoron Trapping}
\label{sec:trapping}

\begin{figure}[H]
  \centering
  \includegraphics[width=0.75\textwidth]{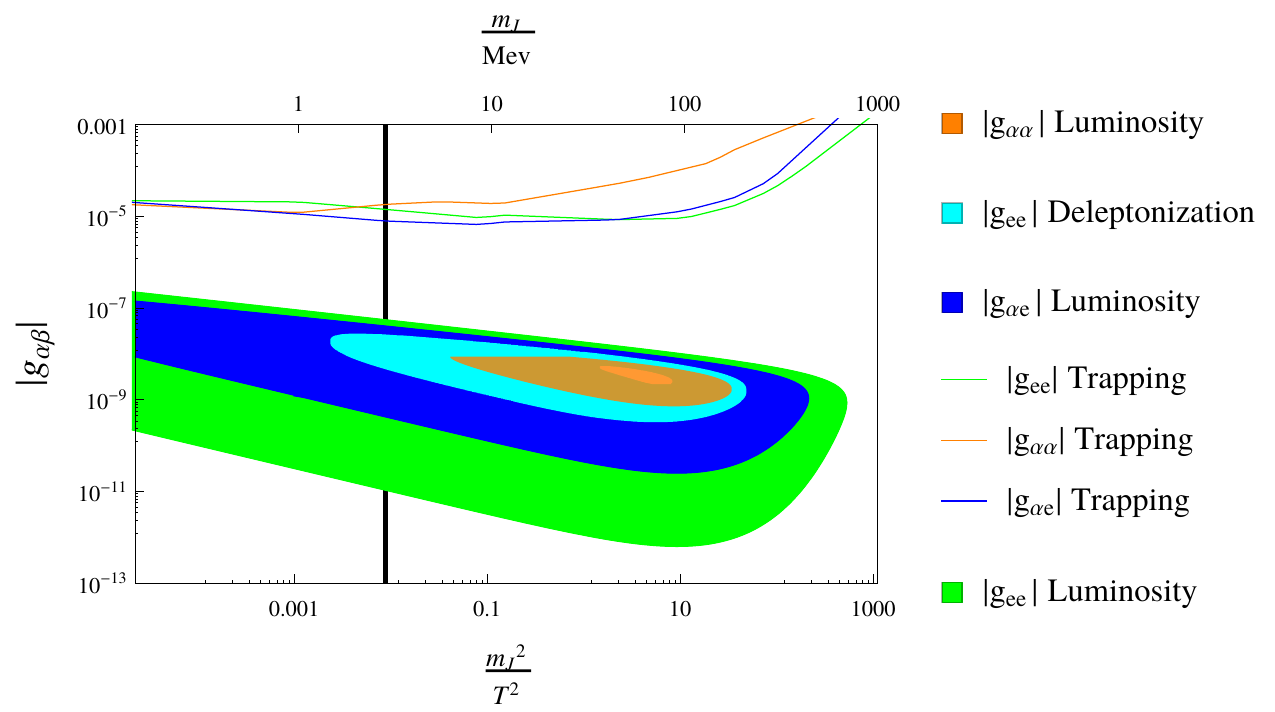}
  \caption{Constraints from SN data. The colored regions are excluded due to luminosity and deleptonization constraints. The region above the lines leads to trapping of Majorons inside of the core. The vertical corresponds to $m_J = \SI{2.8}{\mega\electronvolt}$, i.e. the mass where the Majoron can account for DM. $T$ is the core temperature and $\alpha = \mu, \tau$. The trapping bounds suffer from numerical instabilities that could not be solved in the available computation time.}
  \label{fig:snconstraints}
\end{figure}

So far, it was assumed that the Majorons produced via $\overset{\scriptscriptstyle(-)}{\nu}_\alpha\overset{\scriptscriptstyle(-)}{\nu}_\beta\to J$ either freely leave the core or decay back to neutrinos.
However, if the coupling between neutrinos and the Majoron is too large, neutrino-Majoron scattering $\overset{\scriptscriptstyle(-)}{\nu} J \to \overset{\scriptscriptstyle(-)}{\nu} J$ can lead to trapping of the Majoron in the core. This has two effects. First of all, trapped Majorons do not lead to an additional energy depletion to be measured on earth.\footnote{We neglect the effect of the volume emission from the ``Majoron sphere'', the sphere in which the Majorons are trapped.} Second, if there is a considerable amount of Majorons in the core, the SN dynamics might change drastically, thus our model would not be valid anymore. Therefore, it is reasonable to assume that the constraints only hold if Majorons do not become trapped. \\
The thermal average of the inverse mean free path is given by \cite{heurtierzhang}
\begin{align}
\bar{l}^{-1}_J(m_J,|{g}_{\alpha\beta}|) &= \frac{\sum_{\alpha, \beta} \int \!\!{d}E_i l^{-1}f_J(E_i)}{\int \!\!{d}E_i f(E_i)} \;,
 \label{eq:trapping}
\end{align}
where
\cite{choisantamaria}
\begin{align}
  l_J^{-1} &= \int \frac{d^3p_\alpha}{(2\pi)^3} \sigma(\overset{\scriptscriptstyle(-)}{\nu}_\alpha J \to \overset{\scriptscriptstyle(-)}{\nu}_\beta J)\;.
\end{align}
For a rough approximation, the influence of the effective potentials ia neglected, i.e. the result is a slight underestimation of the mean free path. However, as $p \gg V_\ab$, we assume that the discrepancy is marginal. Moreover, the Majoron distribution function is approximated as the convolution of two neutrino distribution functions,
\begin{align}
 f_J(E_i) &= \frac{T e^{\frac{\mu_a + \mu_b}{T}}}{e^{\frac{\mu_a + \mu_b}{T}} - e^{\frac{E_i}{T}} } \log{\left( \frac{e^{\frac{E_i}{T}}  \inbr{ 1+ e^{\frac{\mu_a}{T}}}\inbr{1+ e^{\frac{\mu_b}{T}}} }{\inbr{e^{\frac{\mu_a}{T}} + e^\frac{E_i}{T} } \inbr{e^{\frac{\mu_b}{T}} + e^\frac{E_i}{T}} }  \right)  }\;.
\end{align}
For the neutrinos, a Fermi-Dirac distribution $f_\alpha = \frac{1}{1+\exp\frac{E_\alpha - \mu_\alpha}{T}}$ is used.\\
We evaluate constraints on $|g_\ab|$ for $\SI{0.1}{\mega\electronvolt} \lesssim m_J \lesssim \SI{1}{\giga\electronvolt}$ from \eqref{eq:trapping} under the condition that Majorons are not trapped, i.e. $\bar l_J > R_C$. The constraints for $|g_\ab|$ are shown in Fig.  \ref{fig:snconstraints}, where the regions above the lines lead to trapping of Majorons. The trapping regions do not intersect the luminosity constraints, i.e. trapping has no impact on our constraints. Note that the ``trapping constraint'' is rather a bound on the validity of our discussion than an actual constraint on the neutrino-Majoron couplings $|g_\ab|$.\\

The combined constraints from SN data, shown in Fig. \ref{fig:snconstraints}, exclude a large region of parameter space. For the neutrino-Majoron couplings $|g_{ee}|$ and $|g_{e\alpha}|\,,\, (\alpha = \mu,\,\tau)$, a range is excluded in which the Majoron could act as DM, i.e. where $m_J \approx \SI{2.8}{\mega\electronvolt}$. The constraints involving nonelectron neutrinos are less extended since their abundance in the SN core is significantly smaller than the abundance of electron neutrinos.
Since an increasing neutrino-Majoron coupling results in a higher Majoron luminosity, one would expect all neutrino-Majoron couplings larger than the respective lower bound of the constraints to be excluded. However, our results show that upper bounds on the constraints exist, i.e. larger values of coupling constants are not excluded due to SN data. This can be traced back to the decay factor $D$ in the calculation of the deleptonization and luminosity constraints: If the neutrino-Majoron coupling is too large, the Majoron decays back to neutrinos before it leaves the core and no exotic energy depletion occurs.
The deleptonization constraints are comparably weak and suffer from the not well understood explosion process. However, including the Majoron in SN simulations could improve the deleptonization constraints. As discussed above and in \cite{heurtierzhang}, the luminosity constraints could allow to exclude a region up to $|g| \approx \num{e-13}$ in the case of the observation of a future nearby SN. The luminosity constraints are discussed in more detail in Sec. \ref{sec:comp}.\\
In \cite{heurtierzhang}, the impact of the effective potentials is neglected. For $m_J \geq \SI{1}{\kilo\electronvolt}$, their impact on the constraints is marginal, i.e. our constraints hardly differ from those presented in \cite{heurtierzhang}. However, for $m_J \leq \SI{1}{\kilo\electronvolt}$, the decay width depends significantly on the effective potentials, which should be included in the constraints presented \cite{heurtierzhang}.\\
For completeness, note that in the case of a massless Majoron, the neutrino decay $\nu \to \bar\nu J$ is kinematically allowed, i.e. the contribution of this process should be included in the calculation of the luminosity and the deleptonization constraints. Therefore, our constraints are only valid for the massive Majoron and in the limit $m_J \to 0$, they do not need to be compatible with constraints on the couplings of a massless Majoron to neutrinos, as calculated for example in \cite{farzan,kachel, tomas}.

\section{Neutrinoless Double Beta Decay with Majoron Emission}

\label{sec:0nubbconstraints}  
Double Beta Decay \cite{zuber, mutoklapdor, rodejohann, doi, doitakasugi, pas, 1208} is a rare nuclear process. It has half-lives of order $\num{e20}$ years or longer and can occur if a single beta decay of the parent nucleus is either energetically forbidden or strongly suppressed due to a large difference in angular momentum. The decay mode first discussed \cite{mayer} is the two-neutrino double beta decay ($2\nu\beta\beta$) 
\begin{align}
  (Z, A) \to (Z+2 ,A) + 2e^- + 2\bar\nu_e\;,
\end{align}
which can be seen as two successive beta decays via virtual intermediate states, where the ordering number $Z$ changes by two units and the atomic mass $A$ remains the same. It is allowed in the SM, independently of the nature of the neutrinos, and is of second order Fermi theory.\\
Another mode is the  neutrinoless double beta decay ($0\nu\beta\beta$) \cite{furry} 
\begin{align}
  (Z, A) \to (Z+2 ,A) + 2e^-\;,
\end{align}
which violates lepton number by two units and is thus forbidden in the SM. 
This decay can only occur if the neutrino is a Majorana particle. Moreover, in order to allow for the helicity matching, the neutrino has to be massive. \\
In the presence of neutrino-Majoron couplings, another possible $0\nu\beta\beta$-mode is the neutrinoless double beta decay with Majoron emission, $0\nu\beta\beta J$ \cite{georgi}, 
\begin{align}
  (Z, A) \to (Z+2 ,A) + 2e^- +J\;.
\end{align}
This mode has been discussed for the case of the massless Majoron (see for example \cite{zwidth, burgess, 95}). The case of the massive Majoron has so far only been discussed in \cite{carone} for the massive vector Majoron and recently in \cite{blum} for the singlet Majoron model.
\subsection{Constraints from $0\nu\beta\beta J$}
In this section, constraints on the effective neutrino-Majoron coupling $ |g_{ee}|$ are derived, based on the nonobservation of $0\nu\beta\beta J$. 
Our analysis follows closely the approach of \cite{blum}, where constraints on $ |g_{ee}|$ for the massive Majoron are derived from limits on $ |g_{ee}|$ for the massless Majoron by taking into account the effect of the mass on the phase space and the signal-to-root-background ratio. \\
Since the $Q$-value of the nucleus determines the Majoron mass which can be probed, data from experiments with high $Q$-values are preferred. The highest $Q$-value comes from \mbox{NEMO-3} using $\ce{^{48}Ca}$ \cite{ca} , however, stronger constraints come from NEMO-3 using $\ce{^{100}Mo}$ \cite{mo}  and $\ce{^{150}Nd}$ \cite{nd} as from EXO-200 using $\ce{^{136}Xe}$ \cite{exo}.\footnote{In \cite{blum}, instead of data from EXO-200, data from KamLAND-Zen using $\ce{^{136}Xe}$ \cite{xe} was included in the analysis. Moreover, in \cite{blum}, data from NEMO-3 using $\ce{^{48}Ca}$ \cite{ca} was not part of the analysis.} Data from experiments using isotopes such as $\ce{^{76}Ge}$ \cite{gerda, heidelbermoscow} and $\ce{^{82}Se}$ \cite{se} are not included in the analysis since they provide comparably weak limits and small $Q$-values. 
The measured $Q$-values and the limits on $|g_{ee}|$ for the massless Majoron can be found in \mbox{Tab. \ref{tab:data}.} 
\begin{table}[H]
	\centering
	\caption{$Q$-values and limits on the neutrino-Majoron coupling $| g_{ee}|$ at $90\%$CL in the case of a massless Majoron.}
	\label{tab:data}
	\begin{tabular}{c c c}
	\toprule
	{Element} & {$Q \:/\: \mathrm{MeV}$} & {$| g_{ee}(m_J = 0) |$} \\
	\midrule 
	$\ce{^{48}Ca}$	&	4.27	&	$(1.0-4.3)\cdot\num{e-4}$ \cite{ca} \\
	$\ce{^{100}Mo}$	&	3.03	&	$(1.6-4.1)\cdot\num{e-5}$ \cite{mo}\\
	$\ce{^{150}Nd}$	&	3.37	&	$(3.3-14.4)\cdot\num{e-5}$ \cite{nd} \\
	$\ce{^{136}Xe}$	&	2.5		&	$(0.8-1.7)\cdot\num{e-5}$ \cite{exo}\\
	$\ce{^{76}Ge}$	&	2.04		&	$(3.4-8.7)\cdot\num{e-5}$ \cite{gerda}\\
	$\ce{^{82}Se}$	&	3.00		&	$(3.2-8)\cdot\num{e-5}$ \cite{se}\\
	\bottomrule
\end{tabular}
\end{table}
The decay rate for $0\nu\beta\beta J$ is given by \cite{doitakasugi}
\begin{align}
  \Gamma^{J} = G^{ J}(Q,Z)| g_{ee}|^2|M^{J}|^2\;,
\end{align}
where $ g_{ee}$ is the effective coupling constant of the Majoron to neutrinos, 
\begin{align}
   g_{ee}  = \sum_{i,j}g_{ij} U_{ei}U_{ej}  \;.
\end{align}
The phase space integral is given by \cite{mutoklapdor}\;,
\begin{align}
\begin{split}
  G^{J} &\propto \int_{m_e}^{E_i-E_f-m_e}\!\!d\epsilon_1F(Z,\epsilon_1)p_1\epsilon_1   \int_{m_e}^{E_i-E_f-\epsilon_1}\!\!d\epsilon_2F(Z,\epsilon_2)p_2\epsilon_2  
  \int\!\! d\epsilon_J p_J \delta(E_i-E_f- \epsilon_J-\epsilon_1-\epsilon_2) \;,
  \label{eq:jansatz}
\end{split}
\end{align}
where $p_J(\epsilon_J)$ is the momentum (energy) of the emitted Majoron, $E_i$ and $E_f$ are the energies of the initial and final state nucleus, respectively, and $\epsilon_{1,2}(p_{1,2})$ are the energies (momenta) of the electrons and $\nu_{1,2}$ are the energies of the neutrinos in the final state.
The Fermi function $F(Z, \epsilon_i)$ takes into account the effect of the Coulomb field of the daughter nucleus $(Z,A)$ on the wave functions of the emitted electrons. \\
Using $Q:= E_i -E_f -2m_e \;$
and the Primakov-Rosen approximation of the Fermi function \cite{primrosen, doitakasugi},  
\begin{align}
  F(Z, \epsilon_i) &= \frac{\epsilon_i}{p_i} \frac{2\pi \alpha Z}{1-\exp(-2\pi\alpha Z)} \;,
\end{align}
the electron sum spectrum can be written as 
\begin{align}
  \frac{dG^{J}}{dT} 
  &\propto \sqrt{(Q -T)^2-m_J^2}\left(30 m_e^4 T+60 m_e^3 T^2+40 m_e^2 T^3+10 m_e T^4+T^5\right) \;,
  \label{eq:spectrum2}
\end{align}
where $T = \epsilon_1 + \epsilon_2 - 2m_e$ is the sum of the kinetic energies of the electrons in the final state.
The decay rate for $2\nu\beta\beta$ can be written as \cite{doitakasugi}
\begin{align}
  \Gamma^{2\nu} = G^{2\nu}(Q,Z)|M^{2\nu}|^2\;,
\end{align}
and the respective electron sum spectrum is given by 
\begin{align}
  \frac{dG^{2\nu}}{dT}&\propto (Q-T)^5\left(30 m_e^4 T+60 m_e^3 T^2+40 m_e^2 T^3+10 m_e T^4+T^5\right) \;.
  \label{eq:spectrum1}\;
\end{align}

\subsubsection{Phase Space Suppression}
\begin{figure}[h]
	\centering
	\includegraphics[width=0.75\textwidth]{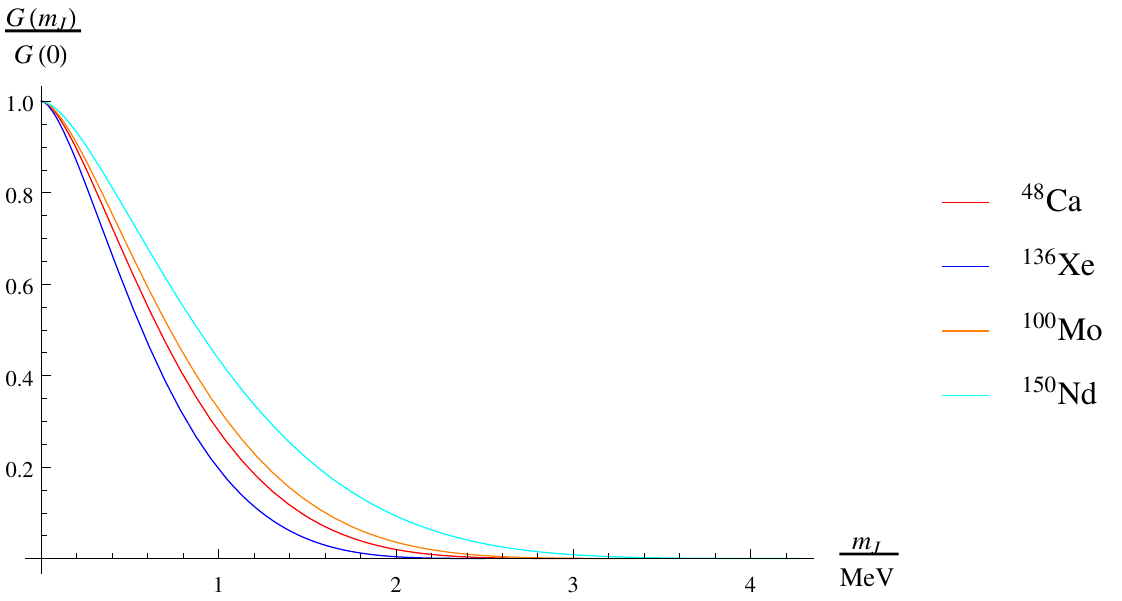}
	\caption{Phase space suppression as a function of $m_J$ in the Primakov-Rosen approximation.}
	\label{fig:pssupp}
\end{figure}

The phase space $G^J(m_J)$ in Eq. \eqref{eq:spectrum2} can be written as    
\begin{align}
	G^J(m_J) = \int_0^{Q-m_J}\!\! \frac{dG^J(m_J)}{dT} dT\;.
\end{align} 
Due to the Majoron mass, the phase space decreases, and the phase space suppression is calculated as $\frac{G(m_J)}{G(0)}$ \cite{blum}.
The normalized ratio $\frac{G(m_J)}{G(0)}$ is plotted in Fig.  \ref{fig:pssupp}. We find $G^J(m_J) \to 0 $ as $m_J \to Q$, thus the decay width is significantly reduced compared to the case of a massless Majoron, resulting in weaker limits on $| g_{ee}(m_J)|$.

\subsubsection{Signal-to-Root-Background Ratio}
\begin{figure}[h]
  \centering
  \includegraphics[width=0.75\textwidth]{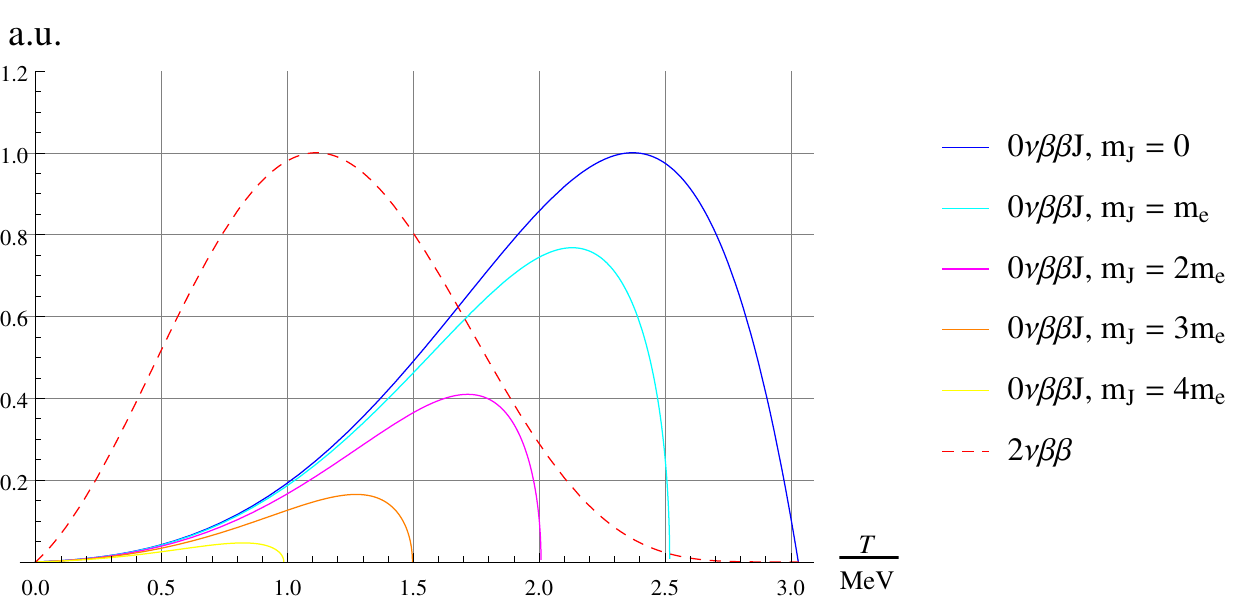}
  \caption{Normalized spectrum of $\ce{^{100}Mo}$ for $2\nu\beta\beta$ and $0\nu\beta\beta J$ for different Majoron masses $m_J$.}
  \label{fig:spectrum_mo}
\end{figure}

\begin{figure}[h]
	\centering
	\includegraphics[width=0.75\textwidth]{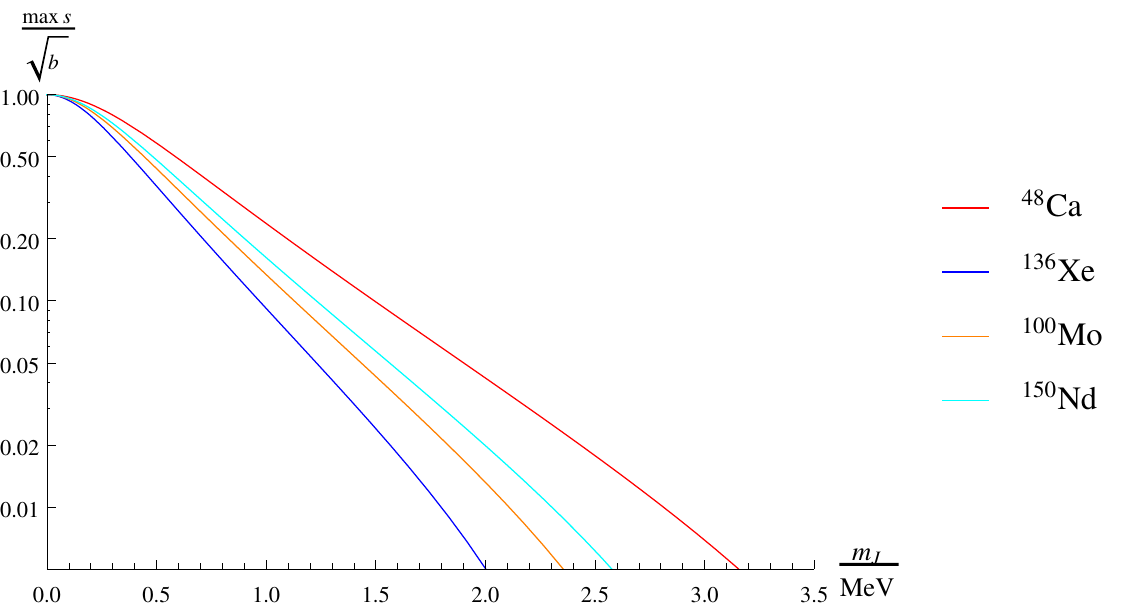}
	\caption{Signal-to-root-background ratio as a function of $m_J$.}
	\label{fig:stob}
\end{figure}

Next, the decrease of the signal-to-root-background ratio is considered, where $0\nu\beta\beta J$ is the signal and $2\nu\beta\beta$ is the background. 
The $\ce{^{100}Mo}$ electron sum spectra are plotted in Fig. \ref{fig:spectrum_mo} for various values of $m_J$. 
Note that the normalizations of $2\nu\beta\beta$ and $0\nu\beta\beta J (m_J=0)$ are arbitrary while the $0\nu\beta\beta J (m_J \neq 0)$ distributions are normalized with respect to $0\nu\beta\beta J (m_J=0)$.\footnote{Our spectra differ from those presented in \cite{blum}. Blum et Al. confirmed a mistake in the relativistic approximation which changes the spectra, however, the effect on the limits presented in \cite{blum} is marginal. } 
With increasing Majoron mass, the relative amplitude and the maximal summed electron energy $T_{max}$ decrease, shifting the spectrum of $0\nu\beta\beta J(m_J\neq 0)$ to the left with respect to the spectrum of $0\nu\beta\beta J(m_J=0)$. Consequently, the overlap with the irreducible SM-$2\nu\beta\beta$ spectrum increases, resulting in a smaller signal ($0\nu\beta\beta J$) to background ($2\nu\beta\beta$) ratio.  \\
For a proper analysis, experiment-dependent sources of background should be incorporated in the analysis. However, as a rough approximation, following \cite{blum}, it is assumed that the impact of the Majoron mass on the spectrum can be taken into account by calculating $\frac{\text{max}[s]}{\sqrt{b}}$, which is normalized with respect to  $\frac{\text{max}[s]}{\sqrt{b}}(m_J = 0)$. As can be seen in \ref{fig:stob}, the signal-to-root-background ratio $\frac{\text{max}[s]}{\sqrt{b}}$  decreases significantly with increasing Majoron mass. 
\subsubsection{Half-life Limit}
\begin{figure}[h]
	\centering
	\includegraphics[width=0.75\textwidth]{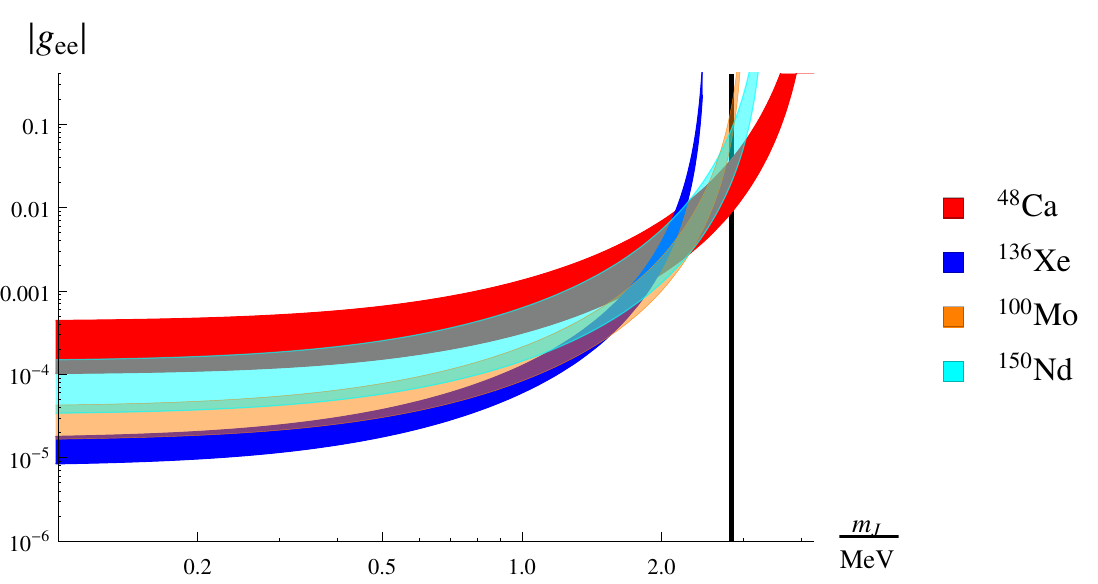}
	\caption{Limits on $| g_{ee}|$ from $0\nu\beta\beta J$ where the region above the bands is excluded by the respective experiment. The widths of the bands represent the uncertainties on $| g_{ee}(0) |$ placed by the collaborations. The vertical line corresponds to $m_J = \SI{2.8}{\mega\electronvolt}$, i.e. the mass where the Majoron can account for DM. }
	\label{fig:0nubbconstraints}
\end{figure}

The decreasing signal-to-root-background ratio $\frac{s}{\sqrt{b}}$ deteriorates the limit on the half life $T_\frac{1}{2}$ as
\begin{align}
	T_\frac{1}{2}(m_J) = \frac{\frac{\text{max}[s]}{\sqrt{b}}(m_J)}{\frac{\text{max}[s]}{\sqrt{b}}(0)}T_\frac{1}{2}(0)\;.
\end{align}
Accordingly, the bound on the effective coupling $|g_{ee}|$ of electron neutrinos to the Majoron can be obtained from
\begin{align}
	| g_{ee}(m_J) | = \sqrt{\frac{G(0)}{G(m_J)}\frac{\frac{\text{max}[s]}{\sqrt{b}}(0)}{\frac{\text{max}[s]}{\sqrt{b}}(m_J)}}| g_{ee}(0) |\;,
\end{align}
where 
\begin{align}
	T^{-1}_\frac{1}{2}(m_J) = G(m_J)|M^J|^2 | g_{ee}(m_J)|^2 \;.
\end{align} 
Here we assumed the nuclear matrix element $M^J$ does not depend on $m_J$. 
The constraints on $|g_{ee}|$ for the kinematically allowed Majoron mass range are plotted in Fig. \ref{fig:0nubbconstraints}. 
We stress that in contrast to Fig. \eqref{fig:snconstraints}, the regions above the bands are excluded in Fig. \ref{fig:0nubbconstraints} by the respective experiment. The widths of the bands represent the uncertainties on $| g_{ee}(0) |$ placed by the collaborations.
Therefore, for $m_J \approx \mathcal{O}(\SI{1}{\mega\electronvolt})$, a large range of neutrino-Majoron couplings $|g_{ee}|$ is excluded. 
Our approach allows us to derive constraints on $|g_{ee}|$ for \textcolor{black}{Majorons in the $\si{\mega\electronvolt}$ mass range} from the constraints on $|g_{ee}(0)|$ provided by the collaborations using $\ce{^{48}Ca}$ and $\ce{^{150}Nd}$. 
We explored the deterioration of the limits based on the decrease of $\frac{\text{max}[s]}{\sqrt b}$, using only $2\nu\beta\beta$ as background. 
For $m_J > 0$, additional sources of background, depending on the respective experiment, should be included, thus we assume to undervalue the uncertainties at $m_J > 0$.

\section{Comparison of the Constraints}
\label{sec:comp}
\begin{figure}[h]
	\centering
	\begin{subfigure}{0.48\textwidth}
		\centering
		\includegraphics[width=\textwidth]{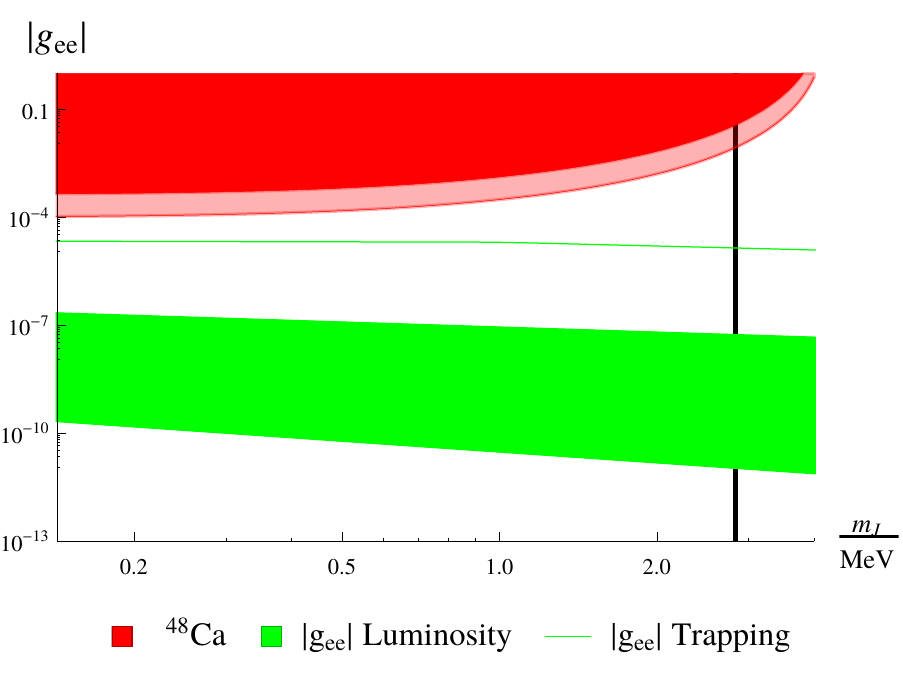}
		\caption{Constraints on $|g_{ee}|$ from SN data and from $0\nu\beta\beta J$ data from NEMO-3 ($\ce{^{48}Ca}$\cite{ca}).}
		\label{fig:comp_ca}
	\end{subfigure}
	\begin{subfigure}{0.48\textwidth}
		\centering
		\includegraphics[width=\textwidth]{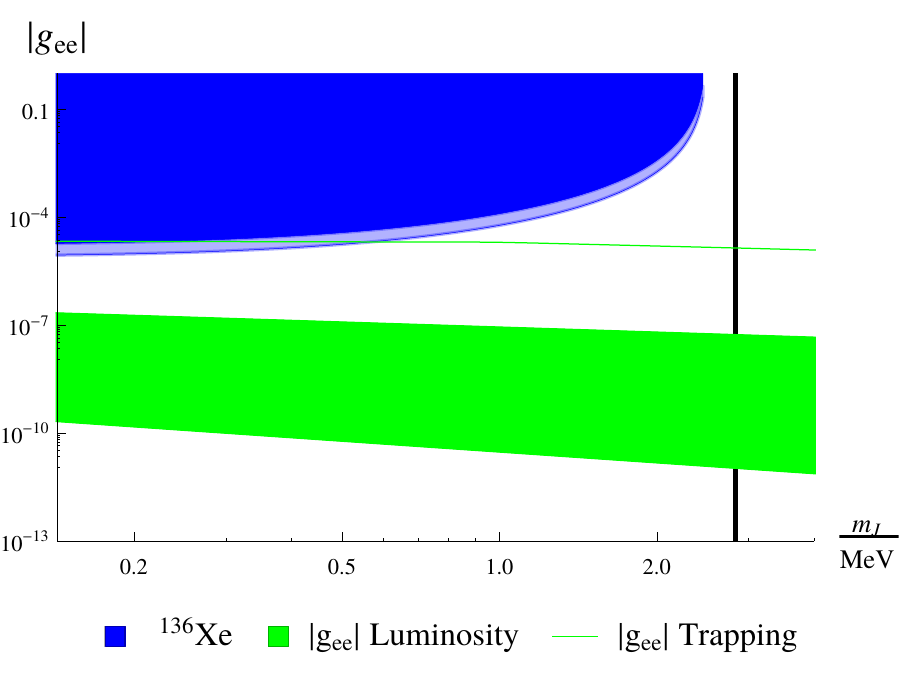}
		\caption{Constraints on $|g_{ee}|$ from SN data and from $0\nu\beta\beta J$ data from EXO-200 ($\ce{^{136}Xe}$\cite{exo}).}
		\label{fig:comp_xe}
	\end{subfigure}
	\caption{Comparison of the constraints on $|g_{ee}|$ from SN data and from $0\nu\beta\beta J$. The colored regions are excluded. The vertical line corresponds to $m_J = \SI{2.8}{\mega\electronvolt}$, i.e. the mass where the Majoron can account for DM. The width of the lighter colored band represents the uncertainties on $|g_{ee}(0)|$ placed by the collaborations.}
	\label{fig:comp}
\end{figure}

In Fig. \ref{fig:comp}, the constraints on $|g_{ee}|$ derived in Sec. \ref{sec:luminosity} from the SN luminosity argument are compared to the $0\nu\beta\beta J$-constraint, discussed in Sec. \ref{sec:0nubbconstraints}. The colored regions are experimentally excluded. For illustrative reasons, only the constraints on $|g_{ee}|$ from NEMO-3 using $\ce{^{48}Ca}$\cite{ca} (Fig. \ref{fig:comp_ca}) and from EXO-200 using $\ce{^{136}Xe}$\cite{exo} (Fig. \ref{fig:comp_xe}) are shown. The width of the lighter colored band represents the uncertainties on $|g_{ee}(0)|$ placed by NEMO-3 \cite{ca} and EXO-200 \cite{exo}, respectively. Since $0\nu\beta\beta J$ allows to derive constraints only on $|g_{ee}|$, we do not show the luminosity or trapping constraints on $|g_{e\alpha}|\,,\, (\alpha = \mu,\,\tau)$ and $|g_{\alpha\alpha}|\,,\, (\alpha = \mu,\,\tau)$. Moreover, the deleptonization constraints on $|g_{ee}|$ are not shown since they do not improve upon the luminosity constraints on $|g_{ee}|$. The vertical line corresponds to $m_J = \SI{2.8}{\mega\electronvolt}$, i.e. the mass where the Majoron can account for DM.\\      
The combination of constraints from SN data and $0\nu\beta\beta J$ data from NEMO-3 \cite{ca} excludes a large range of neutrino-Majoron couplings $|g_{ee}|$ for $\SI{0.1}{\mega\electronvolt} \lesssim m_J \lesssim \SI{5}{\mega\electronvolt}$ and therefore, the constraints are also viable for $m_J \approx \mathcal{O}(\SI{1}{\mega\electronvolt})$, i.e. they provide constraints on Majoron DM produced via freeze-in considered in this work, \textcolor{black}{even though the constraints do not improve upon the already existing constraints from DM stability and CMB anisotropy spectrum.} \textcolor{black}{For the sake of completeness, we want to point to the existence of constraints on neutrino couplings to a massive scalar from Big Bang nucleosynthesis \cite{bbn}.These bounds apply independent of the neutrino flavour for sub-MeV scalar masses and constrain the coupling to be smaller than $\mathcal{O}(10^{-8})$.} Note that the value \textcolor{black}{of the DM mass} is an approximation and an upper bound, assuming the Majoron is the only DM particle. Moreover, other mechanisms to generate a DM relic denstiy with Majorons exist. \\   
Data from EXO-200 using $\ce{^{136}Xe}$\cite{exo} provides the strongest constraints regarding $0\nu\beta\beta J$. Therefore, in the case of a massive Majoron, $0\nu\beta\beta J$ data excludes Majoron trapping up to a Majoron mass of $m_J \approx \SI{0.3}{\mega\electronvolt}$. \\
The $0\nu\beta\beta J$ half-life limit $T_\frac{1}{2}$ depends on the sensitivity of the double beta decay experiment, which is expected to be increased in future experiments. Nonobservation of $0\nu\beta\beta J$  and the sensitivity improvement in the future would translate to increasing half-life limits. Therefore, future $0\nu\beta\beta J$ experiments could probe smaller couplings $|g_{ee}|$, leading to more stringent limits. It will be particularly interesting if the sensitivity of future $0\nu\beta\beta J$ experiments would allow to close the gap between the luminosity constraints on $|g_{ee}|$. Moreover, we only gave a crude estimate on the varying signal-to-root-background ratio for a massive Majoron. An analysis of the effect of a massive Majoron on the signal-to-root-background ratio performed by the respective collaborations could significantly improve the constraints on $|g_{ee}|$.  \\
Our calculation of the luminosity constraints relies on the poor experimental data from SN1987A. 
The detection of a future SN at a distance of order of $\SI{1}{\kilo\parsec}$ would lead to a significant improvement. In \cite{heurtierzhang}, the estimated numbers of detected neutrino events in the Super-Kamiokande and IceCube experiments are $\num{e5}$ and $\num{e8}$, respectively, which would allow us to probe couplings $|g_\ab|$ down to $\num{e-13}$. Consequently, a detected future SN and improved $0\nu\beta\beta J$ experiments could exclude neutrino-Majoron couplings $|g_{ee}|$ down to $\num{e-13}$ for $m_J \approx \mathcal{O}(\SI{1}{\mega\electronvolt})$.

\section{Conclusion}
\label{sec:conclusion}
In this work, we considered a singlet Majoron model, where a pseudo-Goldstone boson, the Majoron, arises at the seesaw-scale due to spontaneous violation of baryon-lepton number $U(1)_{B-L}$.
The couplings of the Majoron to the SM fermions are highly suppressed, rendering it stable on cosmological time scales and thus allowing the Majoron to be a DM candidate. If the Majoron has a mass $m_J \approx \SI{2.8}{\mega\electronvolt}$, the observed DM relic density can be produced by means of the Majoron via the freeze-in mechanism. In this work, constraints on neutrino-Majoron couplings for a Majoron with $m_J \approx \mathcal{O}(\SI{1}{\mega\electronvolt})$ from SN data and neutrinoless double beta decay have been discussed.\\
Neutrinos play an important role in the dynamics of SN explosions, allowing us to derive two bounds on neutrino-Majoron couplings from SN cooling. First, the energy loss of the SN core due to the Majoron emission has to be small compared to the energy emission by neutrinos in order to explain the neutrino signal observed from SN1987A. Second, a large depletion of electron lepton number during the infall stage can prevent a successful explosion. For a $m_J \approx \SI{0.1}{\mega\electronvolt}- \SI{1}{\giga\electronvolt}$, a large region of neutrino-Majoron couplings is excluded (see Fig.  \ref{fig:snconstraints}).
Additionally, we find that Majoron trapping does not affect our constraints. 
We stress that including the Majoron in SN simulations or the observation of a nearby SN explosion could improve or reinforce our constraints. \\
Moreover the couplings of neutrinos to the Majoron can allow for neutrinoless double beta decay with Majoron emission. If kinematically allowed, i.e. $m_J < Q$, the nonobservation of $0\nu\beta\beta J$ translates to constraints on $|g_{ee}|$ (see Fig.  \ref{fig:0nubbconstraints}). The analysis was based on the depletion of the phase space factor $G(m_J)$ and the reduction of the signal-to-root background ratio, $\frac{s}{\sqrt{b}}$, due to the Majoron mass $m_J$. We stress that we did not properly include the background for $m_J > 0$ and urge the collaborations to explore the limits on the massive Majoron model in more detail. Future $0\nu\beta\beta J$ experiments with an improved sensitivity could exclude larger regions of neutrino-Majoron couplings and would therefore strengthen the constraints on the Majoron model. If the increasing sensitivity would allow to close the gap between $0\nu\beta\beta J$ constraints and SN constraints, the electron neutrino-Majoron would be strongly constrained, $|g_{ee}| \lesssim \num{e-11}$ for $m_J \approx \mathcal{O}(\SI{1}{\mega\electronvolt})$. If a neutrino signal of a future galactic SN would be observed, even stronger limits could be obtained, $|g_{ee}| \lesssim \num{e-13}$ for $m_J \approx \mathcal{O}(\SI{1}{\mega\electronvolt})$. \\
\textcolor{black}{The constraints derived from SN data and neutrinoless double beta decay are not yet in the scope of the constraints from DM stability and CMB anisotropy spectrum, i.e. $g < 10^{-20}$. However, the gap between the constraint from Majoron DM and astrophysical and laboratory constraints encourages future experimental efforts and complements the already existing constraints. }\\
To summarize, the constraints from SN cooling and neutrinoless double beta decay with Majoron emission exclude a large space of couplings \textcolor{black}{of a Majoron with a mass in the $\si{MeV}$ range to neutrinos.}
\section*{Acknowledgments}
We thank Kfir Blum for a helpfull discussion regarding $0\nu\beta\beta J$.

\appendix
\section{Appendix}
\subsection{Neutrino-Majoron Interactions in Matter}
\label{app:int}
We write the total Hamiltonian in matter as \cite{giunti}
\begin{align}
  H = H_0 + H_{med}\;,
  \label{eq:matterham}
\end{align}
where $H_0$ is the vacuum Hamiltonian obeying 
\begin{align}
  H_0 \ket{\nu_i} = E_i \ket{\nu_i}
\end{align}
with energy eigenvalues 
\begin{align}
  E_i \approx p  + \frac{m_i^2}{2p}\;,\qquad p:= |\vec p|\;
  \label{eq:energyev}
\end{align}
in the relativistic approximation.
Moreover, $H_{med}$ takes the interaction with the medium into account. The flavor states are eigenstates of the medium Hamiltonian,
\begin{align}
  H_{med} = V_\alpha^{(h)}\ket{\nu_\alpha^{(h)}}\;,
\end{align}
with matter potentials $V_\alpha^{(h)}$, defined in \eqref{eq:pote} and \eqref{eq:potx}.
Neutrino flavor eigenstates $\ket{\nu_\alpha}$ are connected to neutrino mass eigenstates $\ket{\nu_i}$ via
\begin{align}
  \ket{\nu_\alpha} = U_{\alpha i}^* \ket{\nu_i} \label{eq:flavormass}
\end{align}
and, assuming neutrinos are of the Majorana type, we adopt the convention to call Majorana neutrinos with negative helicity neutrinos and Majorana neutrinos with positive helicity antineutrinos: 
\begin{align}
  \ket{\nu^{(h)}} = \begin{cases} \ket{\nu}\,,\;\qquad &h = -1\;, \\  \ket{\bar\nu}\,, &h = +1 \;.  \end{cases} 
\end{align} 
In the mass basis, the Schr\"odinger equation can be written as \cite{kachel, tomas}
\begin{align}
  \mathrm{i}\partial_t \ket{\nu_i^{(h)}} &= \underbrace{(E_i\delta_{ij} + U_{\alpha i}{V}_{\alpha}U_{\alpha j}^* )}_{=:\tilde{H}^m_{ij}}\ket{\nu_j^{(h)}}\;,
  \label{eq:schromatter}
\end{align}
where in matrix form, $\tilde H^m = E + UVU^\dagger$
is nondiagonal, i.e. $\ket{\nu_i}$ is indeed not an eigenstate of the Hamiltonian in matter \eqref{eq:matterham}. 
Introducing a matrix $\tilde U\!\inbr{\theta^{(h)}}$ that diagonalizes $\tilde H^m$, 
  $ \tilde{H}^m_\text{diag} = \tilde{U}^\dagger(\theta^{(h)})\tilde{H}^m\tilde{U}(\theta^{(h)})$,
 results in the relation 
  $\ket{ \tilde{\nu}_i^{(h)}} = \tilde{U}_{ij}(\theta^{(h)})\ket{\nu_j^{(h)}}$
 between medium states $\ket{\tilde\nu_i^{(h)}}$ and mass states $\ket{\nu_j^{(h)}}$, where $\theta^{(h)}$ is the effective mixing angle in matter.\\
In the flavor basis,  the Schr\"odinger equation is given by 
\begin{align}
  \mathrm{i}\partial_t\ket{\nu_\alpha^{(h)}} = \underbrace{(U_{\alpha i}^* E_iU_{i\beta} + {V}_{\alpha}\delta_{\alpha\beta})}_{=: \tilde{H}^w_\ab}\ket{\nu_\beta^{(h)}}\;, 
\end{align}
and in the ultrarelativistic approximation, the relations 
$p \gg \frac{m_i^2}{2p}$ and $V_\alpha^{(h)}| \gg \frac{m_i^2}{2p}$,
hold. Using \eqref{eq:energyev}, the medium Hamiltonian in the weak basis is approximately diagonal, 
\begin{align}
  \tilde{H}^w\ket{\nu_\alpha^{(h)}} &\approx (p + V_\alpha^{(h)}) \ket{\nu_\alpha^{(h)}} \;,
\end{align}
with medium energy eigenvalues
\begin{align}
  E^{(h)} = p+V_\alpha^{(h)}\;.
  \label{eq:meden}
\end{align}
Therefore, the weak states can be approximated as the medium eigenstates\footnote{Since the potential $V$ commutes with $U_{23}$, another choice of medium eigenstate would be $\ket{\tilde\nu_i} \approx U_{23}\ket{\nu_\alpha}$, with the same medium energy eigenvalues \eqref{eq:meden}. The rotation around $U_{23}$ results in easier expressions for the mixing matrix in medium $\tilde U(\theta^{(h)})$. However, a relation between medium and flavor eigenstates is sufficient for our discussion, thus we will stick to the approximation $\ket{\tilde \nu_i}~\approx~\ket{\nu_\alpha}$. },
\begin{align}
  \ket{\tilde \nu_i}\approx\ket{\nu_\alpha}\;.
  \label{eq:massmediumes}
\end{align}
In order to discuss the impact of a background medium on the neutrino-Majoron couplings, the Hamiltonian \eqref{eq:matterham} is extended by a term  
\begin{align}
 H_J = \sum_{i,j}\sum_{h_i,h_j} g_{ij}\bar\nu_i^{(h_i)}\gamma_5\nu_j^{(h_j)}J,
\end{align}
which takes the neutrino-Majoron interactions into account.
In vacuum, the Majoron coupling to the mass eigenstates is diagonal, i.e. $g_{ij} = g_{ij}\delta_{ij}$. Inserting \eqref{eq:flavormass} yields the nondiagonal coupling matrix in the medium basis  
\begin{align}
   \tilde g_{fm} = g_\ab = U_{\alpha i} g_{ij} U_{\beta j}\;.
\end{align}
Thus, in medium, the Majoron couples to the neutrino flavor eigenstates.

\nocite{*}
\bibliography{references}
\bibliographystyle{h-physrev}

\end{document}